\documentclass[jounary,article,accept,pdftex,moreauthors]{mdpi} 
\preto{\abstractkeywords}{\nolinenumbers} 

\firstpage{1} 
\pubvolume{1}
\issuenum{1}
\articlenumber{0}
\pubyear{2022}
\copyrightyear{2022}
\datereceived{} 
\dateaccepted{} 
\datepublished{} 
\hreflink{https://doi.org/} 

\Title{IBM quantum platforms: a quantum battery perspective}

\TitleCitation{Title}

\Author{Giulia Gemme $^{1}$, Michele Grossi $^{2}$\orcidB{}, Dario Ferraro $^{1,3*}$\orcidA{}, Sofia Vallecorsa $^{2}$ and Maura Sassetti $^{1,3}$}

\AuthorNames{Firstname Lastname, Firstname Lastname and Firstname Lastname}

\AuthorCitation{Lastname, F.; Lastname, F.; Lastname, F.}

\address{%
$^{1}$ \quad Dipartimento di Fisica, Università di Genova, Via Dodecaneso 33, 16146, Genova\\
$^{2}$ \quad CERN, 1 Esplanade des Particules, Geneva CH–1211, Switzerland\\
$^{3}$ \quad CNR-SPIN, Via Dodecaneso 33, 16146, Genova}

\corres{Correspondence: ferraro@fisica.unige.it}

\abstract{We characterize for the first time the performances of IBM quantum chips as quantum batteries, specifically addressing the single-qubit Armonk processor. By exploiting the Pulse access enabled to some of the IBM Quantum processors via the Qiskit package, we  investigate advantages and limitations of different profiles for classical drives used to charge these miniaturized batteries, establishing the optimal compromise between charging time and stored energy. Moreover, we consider the role played by various possible initial conditions on the functioning of the quantum batteries. 
As main result of our analysis, we observe that unavoidable errors occurring in the initialization phase of the qubit, which can be detrimental for quantum computing applications, only marginally affects energy transfer and storage. This can lead counter-intuitively to improvements of the performances. This is a strong indication of the fact that IBM quantum devices are already in the proper range of parameters to be considered as good and stable quantum batteries, comparable to state of the art devices recently discussed in literature.}

\keyword{Quantum batteries; Time-dependent quantum transport; Quantum technology} 

\begin{document}


\section{Introduction}

Quantum Batteries (QBs) are miniaturized devices exploiting purely non-classical features in order to outperform their classical counterparts in terms of energy storage, charging power and work extraction~\cite{Campaioli18,Bhattacharjee21}. They recently emerged as a fast growing and very active field of research in the domain of quantum technologies. These investigations represent a radical change of perspective in the framework of energy manipulation with respect to the electrochemical principles developed in the Eighteenth and Nineteenth centuries which are still at the core of nowadays technology~\cite{Vincent_Book, Dell_Book}. 

Starting from seminal ideas developed in Ref.~\cite{Alicki13}, the theoretical investigations in this domain have been at first characterized by the influence of theorems mediated by quantum information~\cite{Binder15,Campaioli17,JuliaFarre20,Gyhm22}. During the years, the studies progressively moved towards more experimentally oriented proposals. They addressed set-ups conveniently designed in such a way to be easily implemented on existing quantum computing platforms such as arrays of artificial atoms~\cite{Le18,Liu19,Rossini20,Rosa20,Crescente20,Carrega20,Santos21,Peng21} and systems for cavity and circuit quantum electrodynamics~\cite{Ferraro18,Ferraro19,Crescente20b,Delmonte21,Dou22}. Very remarkably, the first experimental evidence of a QB has been reported less than one year ago in a system where fluorescent organic molecules play the role of two-level systems embedded in a microcavity~\cite{Quach22}. Even more recently QBs realized with transmon qubits~\cite{Hu21} and quantum dots~\cite{Wenniger22} have been reported, further testifying the great ferment around this topic.

The majority of the considered approaches for QBs are based on two-level systems (qubits) promoted from the ground to the excited state by means of the action of another system playing the role of a charger~\cite{Andolina18,Qi21,Crescente22}. This latter element can be genuinely quantum, such as photons trapped into a cavity~\cite{Ferraro18}, or more simply a classical time dependent drive directly applied to the qubit~\cite{Zhang19,Crescente20,Chen20}. The efficiency of this kind of charging processes is characterized in terms of figures of merits such as the energy stored into the QB, the charging time required to reach its maximum and the average charging power, which is the energy stored in a given time~\cite{Binder15}. \\
IBM quantum devices~\cite{Corcoles20} offer the unique opportunity to simulate quantum systems under controlled conditions, leading to an exponentially increasing number of scientific paper covering various branches of research.
The following are some examples that go from quantum chemistry and material sciences
\cite{Cao_2019,Guimar_2020}, to the analysis of molecular magnetic clusters and spin-spin dynamical correlation functions \cite{Chiesa_2019,Crippa_2021}, 
up to quantum field theories \cite{Fillion_Gourdeau_2017,Klco_2020}, 
high energy physics \cite{bauer2021quantum,grossi_2022} and dark matter \cite{Cervia2020ExactlySM}.
Lots of different publications are defined within real use cases belonging to different industries, including finance, material science and optimization \cite{Pistoia22,Gao2021,Moll_2018}.

An important step forward in the direction of addressing quantum dynamics has been represented by the implementation of the Pulse tool included in the Qiskit package~\cite{Alexander20}, which opened the way to the possibility to reach an unprecedented level of control over the form and the relevant parameters of a classical drive applied to quantum systems. 

Aim of this paper is to realize the first simulation of a classically driven QB applying different controlled pulses on an IBM quantum device. We will focus on the simplest possible machine, the Armonk quantum processor, made by a single transmon qubit.
After a proper calibration of the data, we will characterize the charging profile of the QB as a function of the time integral of the envelop function of the pulse. We will determine the constraints on the form of the pulse in order to obtain an universal charging curve and we will establish the minimum reachable charging time. Without implementing any ad hoc optimization procedure, we observe performances in terms of charging time and stored energy compatible with state of the art experiments in the domain~\cite{Hu21}. Moreover, we observe that initialization errors ubiquitously present in the Noisy Intermediate-Scale Quantum devices~\cite{Preskill18}, which can be detrimental in a quantum computation perspective, can lead to an improvement of the performances of these set-ups as a QB. 

The present paper is organized as follows. In Section \ref{Model} we investigate the model of a qubit coupled with a time dependent drive as a proper description of the Armonk IBM quantum processor subject to the action of the Pulse tool. The calibration of the system and the analysis of the data provided by the IBM interface are reported in Section \ref{Calibration}. The discussions concerning the universality of the charging curve, the possible technical constraints on the suitable forms of the pulses, the achievable minimal charging time and the role played by different initial conditions are reported in Section \ref{Results}. Finally Section \ref{Conclusions} is devoted to the conclusions.

\section{Model}
\label{Model}

We consider a superconducting qubit in the transmon regime~\cite{Koch07}. In the working conditions investigated in this paper, it can be seen as an effective two-level system driven in time by a classical pulse. Its Hamiltonian reads ($\hbar=1$)

\begin{eqnarray}
H&=&H_{QB}+H_{C}\\
&=&\frac{\Delta}{2}\left(1- \sigma_{z}\right)+g f(t) \cos{\left(\omega t\right)}\sigma_{x}
\label{H}
\end{eqnarray}
where the first term ($H_{QB}$) represents the free Hamiltonian of a QB with a level spacing $\Delta$ between the ground state $|0\rangle$ and the excited state $|1\rangle$, while the second term ($H_{C}$) describes the classical charging of the QB itself due to the application of a time dependent drive compatible with the Qiskit Pulse tool~\cite{Alexander20}.  
In the above equation, with $\sigma_{x,z}$ we indicate the Pauli matrices along the $x$ and $z$ direction respectively. Notice that one has
\begin{eqnarray}
\sigma_{z}|0\rangle&=&|0\rangle\\
\sigma_{z}|1\rangle&=&-|1\rangle.
\end{eqnarray}
With $f(t)$ we donote a time dependent adimensional envelop function with maximum amplitude equal to one, whose form will be specified in the following. It is further modulated by a cosine function with frequency $\omega$. Moreover, $g$ represents the coupling between the QB and the classical drive. Aim of this Section is to study the dynamics of this system starting from the generical initial wave-function at time $t=0$
\begin{equation}
|\Psi(0)\rangle=\alpha |0\rangle+\beta |1\rangle
\end{equation}
with $\alpha$ and $\beta$ complex parameters satisfying $|\alpha|^{2}+|\beta|^{2}=1$. \\
Typically, the IBM quantum machines are built in such a way that $\Delta\gg g$. In particular, for the Armonk quantum processor used in this work one has $\Delta\approx 31.238\,\,\mathrm{GHz}$ and $g\approx0.105\,\,\mathrm{GHz}$. Under this condition, for arbitrary values of driving frequency $\omega$ it is not possible to achieve a transition from $|0\rangle$ to $|1\rangle$, namely a charging of the QB~\cite{Crescente20}. \\
This issue can be overcome by properly tuning the systems in the perfectly resonant case $\Delta=\omega$. The reason for the peculiarity of this case can be better appreciated moving to a rotating frame, namely by applying the time dependent rotation 
\begin{equation}
S(t)=e^{-i \frac{\Delta}{2} t \sigma_{z}}    
\end{equation}
to the Hamiltonian in Eq. (\ref{H}) in such a way to obtain 
\begin{equation}
H'=S H S^{\dagger}-i S \frac{ d S^{\dagger}}{dt}, 
\label{H_prime_general}
\end{equation}
where the time dependence of the operators has been omitted for notational convenience. 
By further considering the rotating wave approximation~\cite{Schleich_Book}, which is very well justified under the conditions of resonance and small coupling we are considering, one can write 
\begin{equation}
H'\approx\frac{g}{2} f(t) \sigma_{x} +\frac{\Delta}{2}, 
\label{H_prime}
\end{equation}
where the constant term plays no role in the dynamics and will be neglected in the following. \\
The above expressions for the rotated Hamiltonian together with the fact that the considered rotation doesn't affect the initial state of the system, namely 
\begin{equation}
|\Psi'(0)\rangle=S(0)|\Psi(0)\rangle=|\Psi(0)\rangle,
\end{equation}
allows to analytically solve the dynamics of the considered qubit. Indeed, one can introduce the eigenstates of the $\sigma_{x}$ operator  
\begin{equation}
    |\pm \rangle= \frac{1}{\sqrt{2}}
\left(|0\rangle\pm |1 \right)
\end{equation}
in such a way that the initial wave-function can be written as 
\begin{eqnarray}
|\Psi'(0)\rangle&=& \left(\frac{\alpha+\beta}{\sqrt{2}} \right)|+\rangle+ \left(\frac{\alpha-\beta}{\sqrt{2}} \right)|-\rangle\\
&=& C_{+}(0)|+\rangle+C_{-}(0)|-\rangle. 
\end{eqnarray}
In this basis the time evolution of the coefficients $C_{\pm}$ satisfies 

\begin{equation}
\frac{d C_{\pm}}{dt}=\mp i \frac{g}{2} f(t) C_{\pm}(t)
\end{equation}
and consequently 
\begin{equation}
 |\Psi'(t)\rangle = C_{+}(t)|+\rangle+C_{-}(t)|-\rangle.    
\end{equation}
with
\begin{equation}
C_{\pm}(t)=C_{\pm}(0)e^{\mp i \frac{g}{2} \int^{t}_{0}f(\tau) d\tau}.
\end{equation}

In the following we will focus on a situation, relevant for the considered quantum simulations, where the state of the system is measured at a given time $t=t_{m}$ such that 

\begin{equation}
C_{\pm}(t_{m})=C_{\pm}(0) e^{\mp i \frac{\theta(t_{m})}{2}}
\end{equation}
with 
\begin{equation}
    \theta(t_{m})= g \int^{t_{m}}_{0} f(\tau) d \tau.
    \label{theta}
\end{equation}
Going back the the original basis one obtains directly 

\begin{eqnarray}
|\Psi(t_{m})\rangle&=& S^{\dagger}(t_{m})|\Psi'(t_{m})\rangle\\
&=& e^{i \varphi(t_{m})}\left[\alpha \cos\frac{\theta(t_{m})}{2}- i \beta \sin\frac{\theta(t_{m})}{2}\right]|0\rangle\nonumber\\
&+&e^{-i \varphi(t_{m})}\left[\beta \cos\frac{\theta(t_{m})}{2}- i \alpha \sin\frac{\theta(t_{m})}{2}\right]|1\rangle
\end{eqnarray}
with 
\begin{equation}
    \varphi(t_{m})=\frac{\Delta}{2} t_{m}.
\end{equation}

According to this analysis, the energy stored into the QB at the measurement time ($t_{m}$) is given by~\cite{Ferraro18, Andolina18}

\begin{eqnarray}
E(t_{m})&=& \frac{\Delta}{2} \langle \Psi(t_{m})|\left( 1-\sigma_{z}\right)|\Psi(t_{m})\rangle\\
&=&\Delta \langle \Psi(t_{m})|1\rangle \langle 1 |\Psi(t_{m})\rangle\\
&=& \Delta | \langle \Psi(t_{m})|1\rangle|^{2}\\
&=& \Delta P_{1}(\theta(t_{m}))
\label{E_m_gen}
\end{eqnarray}
where the last line indicates the probability to find the QB in the excited state ($|1\rangle$) when the measurement is carried out. In the following we will omit the dependence on $t_{m}$ for notation convenience. \\
Taking into account the reparametrization 
\begin{eqnarray}
\alpha&=&\sqrt{a}\\
\beta&=& \sqrt{1-a} e^{-i \phi}, 
\end{eqnarray}
with $a$ and $\phi$ real numbers, one can finally write 
\begin{equation}
E(a, \phi, \theta)=\Delta  \left[a \sin^{2}{\frac{\theta}{2}} +2 \sqrt{a} \sqrt{1-a} \sin\phi\sin{\frac{\theta}{2}}\cos{\frac{\theta}{2}}+(1-a) \cos^{2}{\frac{\theta}{2}}\right],  
\label{E_m_par}
\end{equation}
where we have made explicit the dependence of the stored energy on the relevant parameters characterizing the initial wave-function of the qubit and the phase associated to the envelop function of the applied pulse. In the following, the results of the simulations carried out as a function of $\theta$ will be fitted by means of Eq. (\ref{E_m_par}) in order to extract the values of the parameters $a$ and $\phi$. This will allow us to reconstruct the initial state of the QB and characterize its performances in terms of stored energy. Notice that a similar analysis has been carried out very recently in an experimental work devoted to the characterization of a semiconducting quantum dot as a QB~\cite{Wenniger22}.

\section{Calibration}
\label{Calibration}

Before entering into the details of the QB charging, we need to discuss the nature of the data returned by the IBM platform and the way to analyze them. The measurement of the state of the qubit after the application of the pulse is done through a readout in the so called dispersive regime~\cite{Krantz19}. Here, an harmonic oscillator (resonator) is weakly coupled to the QB. The system is then described by the Jaynes-Cummings Hamiltonian~\cite{Schleich_Book}
\begin{equation}
H_{RO}=H_{QB}+\omega_{r} b^{\dagger}b+\lambda \left( b^{\dagger}\sigma_{-} + b\sigma_{+}\right)    
\label{H_RO}
\end{equation}
where $\lambda$ indicates the strength of the matter-radiation dipolar coupling and with $b^{\dagger}$/$b$ and $\sigma_{\pm}$ ladder operators for the harmonic oscillator and the effective spin associated to the QB, respectively. When the two part of the system are kept far from resonance, namely under the condition $\lambda \ll |\Delta-\omega_{r}|$, it is possible to perform a Schrieffer-Wolff transformation and to expand Eq. (\ref{H_RO}) up to the second order in the coupling, obtaining the effective Hamiltonian~\cite{Schleich_Book}

\begin{equation}
H_{eff}=H_{QB}+\omega_{r} b^{\dagger}b-\frac{2 \lambda^{2}}{|\Delta-\omega_{r}|} \sigma_{z} \left(b^{\dagger} b+\frac{1}{2}\right).
\end{equation}
This leads to a shift in the frequency of the oscillator which depends on the state of the QB. \\
Under these conditions, a monochromatic microwave with frequency $\Omega$ applied to the resonator is modified in such a way that  
\begin{equation}
    \cos{\Omega t}\rightarrow  \mathcal{A} \cos \left(\Omega t+\chi\right).
\end{equation}
Namely it acquires a different amplitude $\mathcal{A}$ and phase $\chi$. These quantities depends on the properties of the resonator and consequently are unambiguously related to the state of the QB. Getting rid of the time dependence of the signal and taking into account the complex representation of the transmitted wave one can write  
\begin{equation}
    \mathcal{A}e^{i \chi}=I+iQ, 
\end{equation}
with $I$ and $Q$ real numbers. 

\begin{figure}[H]
\includegraphics[width=7.0 cm]{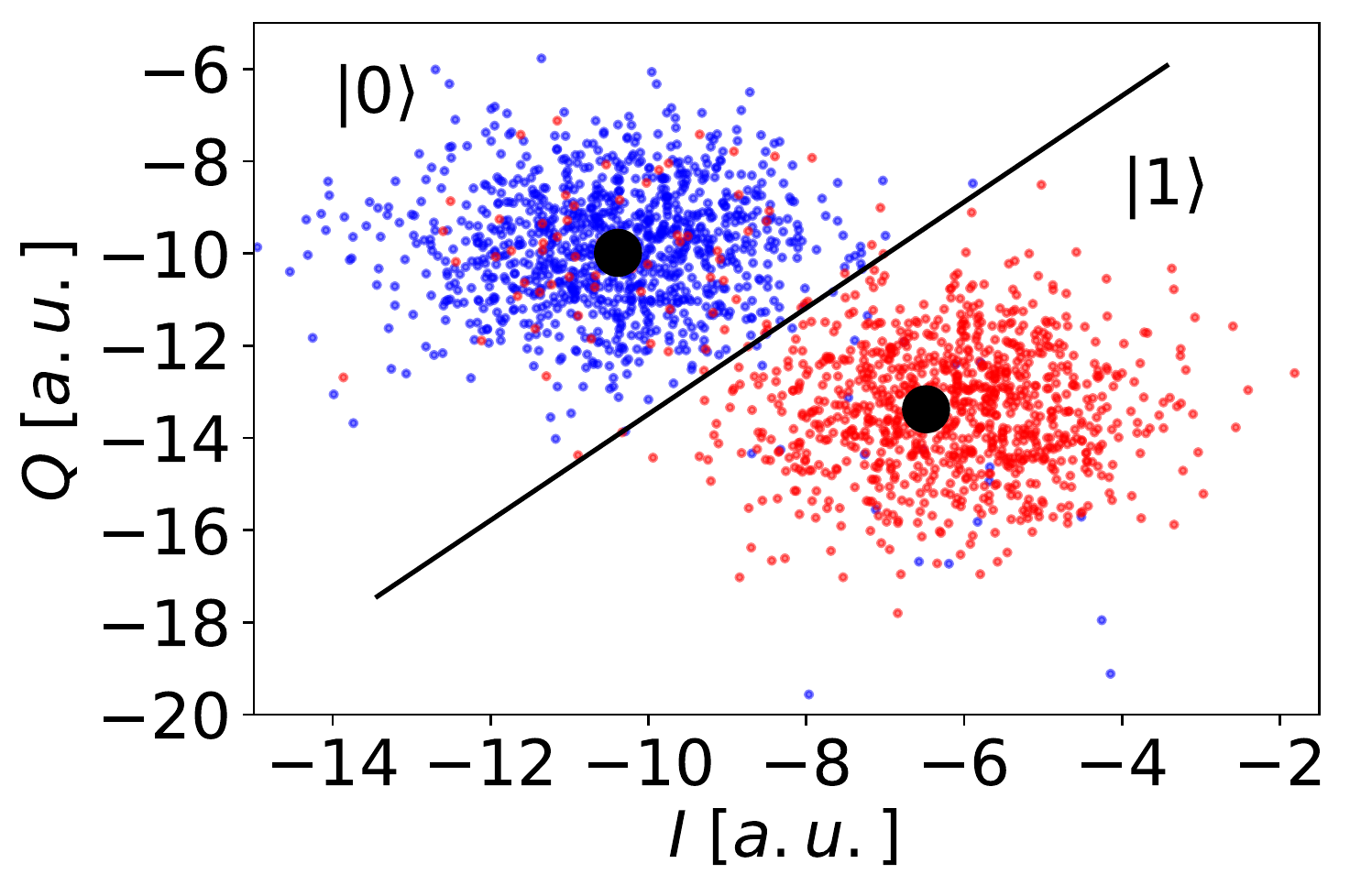}
\includegraphics[width=7.0 cm]{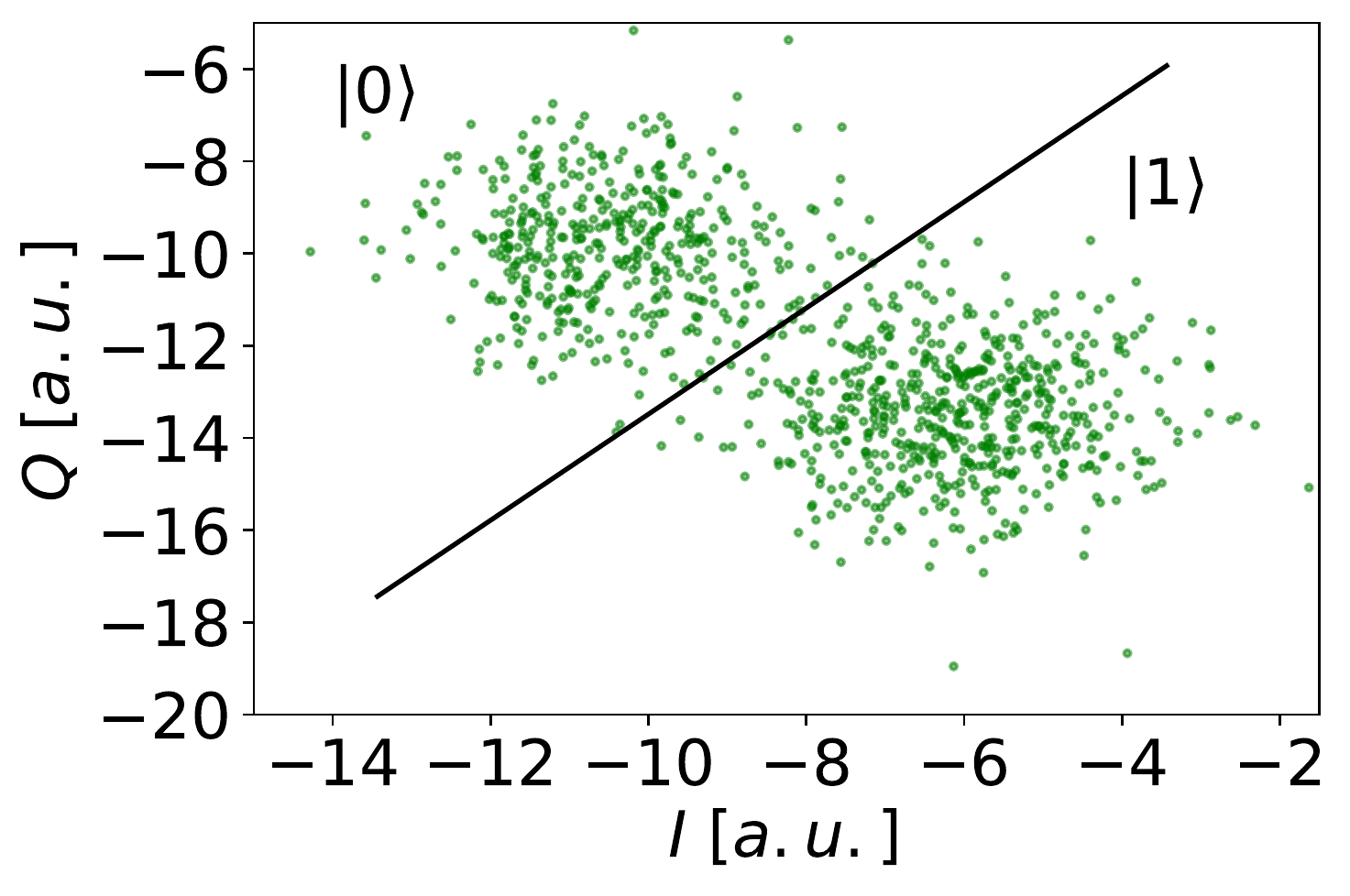}
\caption{(Color online). Left panel. Example of data distribution associated to the measurements of the state $|0\rangle$ (blue dots) and $|1\rangle$ (red dots) in the $(I, Q)$ plane (in arbitrary units). Big black dots indicate the centers of the two distributions, while the black line separates them. Notice that, despite the distortion due to the aspect ratio of the plot, this line is perpendicular to the segment connecting the two centers. The efficiency of the considered separation is roughly $97,4\%$ for the ground state and $92,7\%$ for the excited state. Right panel. Example of a distribution associated to the measurement of a state with $\theta=\pi/2$ (green dots). The distribution presents now two separated lobes, one in each sector. The energy stored in the system is related to the fraction of the dots composing the lobe in the $|1\rangle$ sector with respect to the total runs. In the present case is given by $P_{1}\approx0.567$. For each state shown in the plot we have considered $1024$ runs.}
\label{fig1}
\end{figure}

Every measurement of the qubit state is then reported as a point in the $(I, Q)$ plane~\cite{Jeffrey14}. In order to accumulate a proper statistics, the machine performs multiple measurements ($1024$ in default settings). They are typically very scattered, requiring a calibration to extract a meaningful information from them. 
In this platform each qubit is initialized to the ground state, so as a first thing we have characterized the ground state $|0\rangle$ of the system measuring it just after the initialization of the machine. Then the excited state $|1\rangle$  has been obtained initializing the system in $|0\rangle$ and  applying a built-in pulse with $\theta=\pi$ (see Eq. (\ref{theta})) of the form 
\begin{equation}
f(t)=\sqrt{\frac{\pi}{2}}\frac{1}{g \sigma} e^{-\frac{\left(t-t_{m}/2 \right)^{2}}{2 \sigma^{2}}}   
\label{calibration_f}
\end{equation} 
with standard deviation $\sigma=t_{m}/8$, being $t_{m}=600\,\,\mathrm{ns}$ the measuring time. A typical distribution of the points in the $(I, Q)$ plane associated to the ground and excited state is reported in the left panel of Fig. (\ref{fig1}). 

In order to discriminate between the two states of the system, we have written a Python routine able to determine the center of the relative distributions (big black dots) and identify the line perpendicular to the segment connecting them and passing through its middle point. This approach implicitly assume that both distributions have the same spread. This line divides the $(I, Q)$ plane into two parts that we can identify as the $|0\rangle$ and the $|1\rangle$ sector respectively. It is worth to note that, due to the intrinsic errors affecting the machine a small fraction (few percent) of the blue dots corresponding to ground state measurement fall into the part of the plane associated to the excited state viceversa. The evaluation of these errors, as well as a reasonable hypothesis of their origin, will be relevant points of the analysis of the results.   \\
According to the picture discussed above, the energy stored in the QB, proportional to the probability for a state to be measured in $|1\rangle$ after the application of a generical pulse with $0\leq \theta \leq \pi$ (see Eq. (\ref{E_m_gen})), can be evaluated directly as the ratio of the number of points falling in the $|1\rangle$ sector with respect to total runs performed by the machine (see the right panel of Fig. \ref{fig1} for an example). 

\section{Results}
\label{Results}

\subsection{Universal charging behavior and technical constraints on the pulses}
In this Section, we consider the charging curve of the Armonk QB for different profiles of the time dependent envelop function $f(t)$. On a very general ground we observe that Eq. (\ref{E_m_par}) doesn't explicitly depend on its functional form, but only on its integral up to the measurement time indicated with $\theta$ (see Eq. (\ref{theta})). Therefore, we expect this universal behaviour to emerge in the measurements. Moreover, we attribute deviations with respect to the ideal curve 
\begin{equation}
E(1, \phi, \theta)=\Delta \sin^{2}\frac{\theta}{2}
\label{E_m_ideal},
\end{equation}
corresponding to the perfect initialization in the ground state, to errors occurring in this phase.

\begin{figure}[H]
\includegraphics[width=14.0 cm]{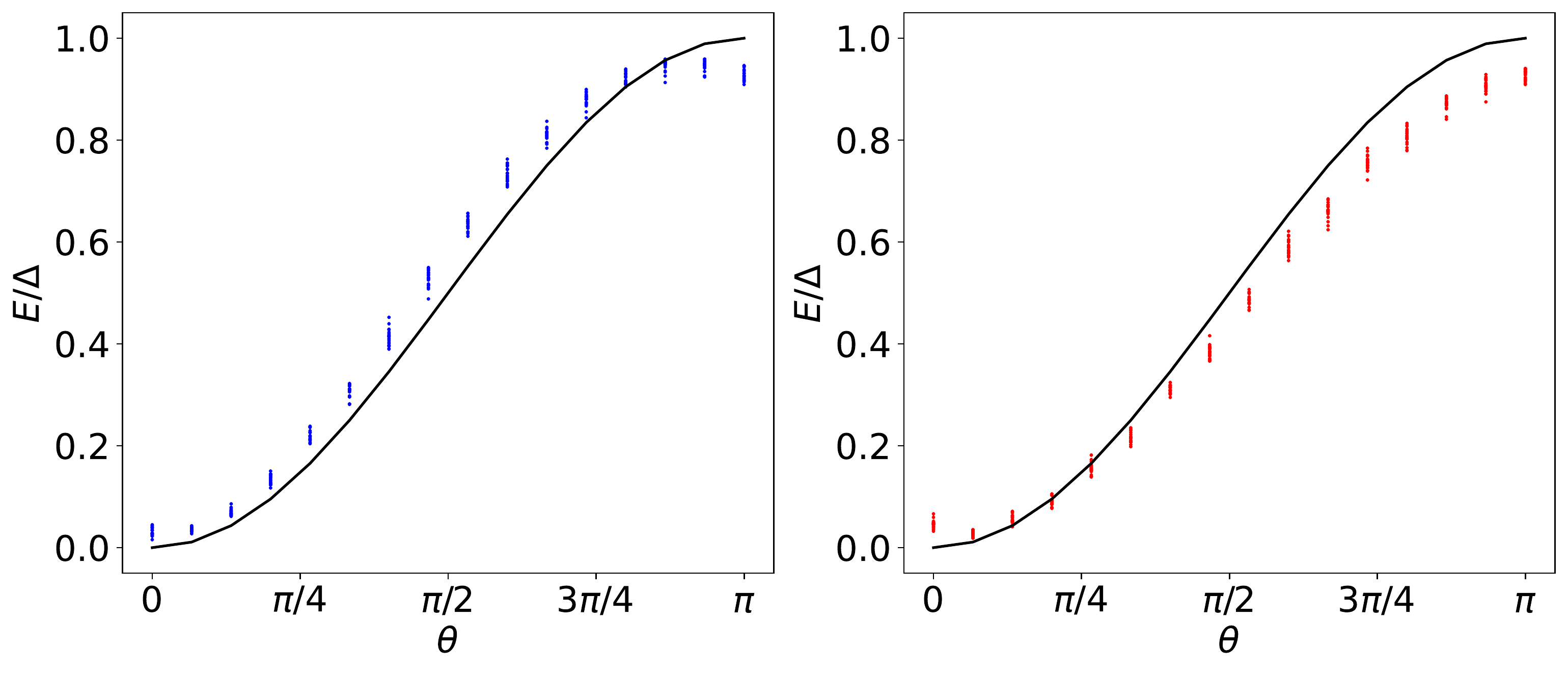}
\caption{(Color online). Energy stored in the QB (in units of $\Delta$) as a function of the area $\theta$. The black curves corresponds to the ideal charging profile with $a=1$ (see Eq. (\ref{E_m_ideal})). Left panel shows the charging achieved using a Gaussian pulse with fixed standard deviation $\sigma=t_{m}/8$, being $t_{m}=600\,\,\mathrm{ns}$ the measurement time, and amplitude $\mathcal{N}=\theta/(\sqrt{2 \pi} g \sigma)$. Right panel shows the charging due to a Gaussian pulse with fixed amplitude equal to $\mathcal{N}=1$ and adjustable standard deviation $\sigma=\theta/(\sqrt{2 \pi} g \mathcal{N})$. The vertical spread of the points is due to the fact that we have considered $20$ measurements for each value of $\theta$.}
\label{fig2}
\end{figure} 

The results concerning Gaussian profiles of the form
\begin{equation}
    f(t)=\mathcal{N}e^{-\frac{\left(t-t_{m}/2 \right)^{2}}{2 \sigma^{2}}},   
\end{equation}
are reported in Fig. \ref{fig2}. Due to the overall constraint given by Eq. (\ref{theta}) the amplitude $\mathcal{N}$ and the standard deviation $\sigma$ cannot be tuned independently. Therefore, we have compared two possible complementary approaches. In the left panel, generalizing what done in Eq. (\ref{calibration_f}), the standard deviation $\sigma$ has been kept fixed to a convenient fraction of the measurement time $t_{m}$ while the amplitude $\mathcal{N}=\theta/(\sqrt{2 \pi} g \sigma)$ (always satisfying $\mathcal{N}<1$) has been varied in order to return the proper value of $\theta$. In the right panel we proceeded in the opposite way, keeping fixed the amplitude to the maximum possible value $\mathcal{N}=1$ and varying the standard deviation according to $\sigma=\theta/(\sqrt{2 \pi} g \mathcal{N})$. It is evident that in both cases the data depart from the ideal behavior of Eq. (\ref{E_m_ideal}). Moreover, despite equivalent from the mathematical point of view, these two approaches lead to different results. This is due the fact that the second approach is affected by a technical problem. For the considered values of $\theta$, the support of $f(t)$ is reasonably different from zero only in a narrow time window. Due to the fact that the Pulse tool discretizes the signals with a minimum time step of $\delta=0.222\,\,\mathrm{ns}$ this peaked envelop function is strongly affected by this and is characterized by a effective area smaller with respect to what expected. Even if this error can be mitigated by considering lower amplitudes $\mathcal{N}$, the first approach revealed more suitable for our purposes with different choices of standard deviation satisfying $\sigma<t_{m}/5$ (not shown) leading to almost superposing distributions of the points. This latter condition is justified by the need of minimizing the error made by cutting the envelop function at a finite measurement time $t_{m}$. While this error is typically under control for fast decaying envelop functions such as the Gaussian considered here, broader profiles like the Lorentzian one require to be properly adjusted in order to compensate for this missing-tail effect and recover the predicted universal behavior. 

The above discussion indicates that a not too narrow and fast decreasing envelop function represents the better choice for a pulse leading to a suitable charging of the considered QB. Once these requirements are properly fulfilled it is also possible to reduce the charging time with respect to the value reported in Fig. \ref{fig2}. This point, together with a fit of the data in order to determine the actual initial state of the system will be investigate in the following. 

\subsection{Best fit of the data and characterization of the QB performances}

\begin{figure}[H]
\includegraphics[width=14.0 cm]{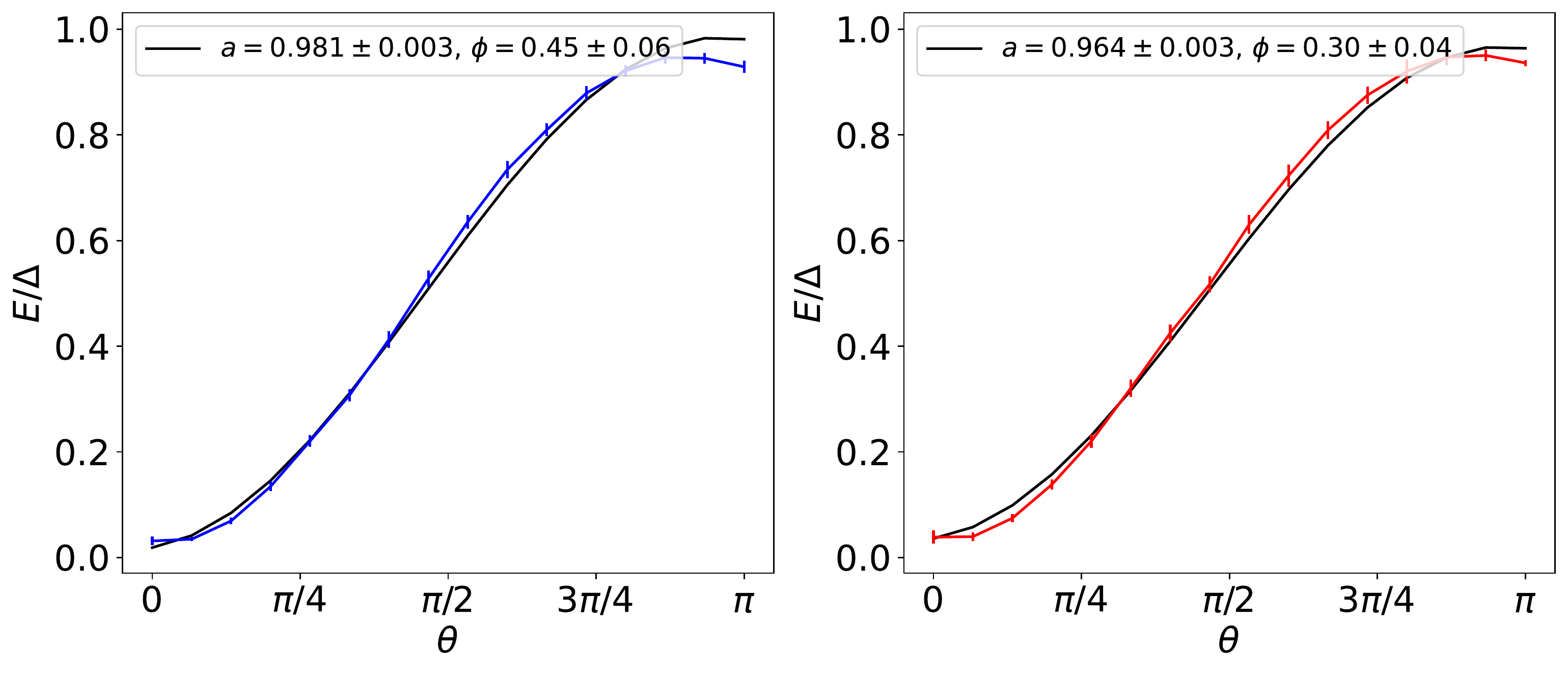}
\caption{(Color online). Best fit of the energy stored into the QB (in units of $\Delta$) as a function of $\theta$ (black curves). Data correspond to Gaussian pulses with $\sigma=t_{m}/8$, being $t_{m}=600\,\,\mathrm{ns}$ (blue curve in the left panel) and $t_{m}=135\,\,\mathrm{ns}$ (red curve in the right panels). Every point is obtained averaging over $20$ different measurements (see Fig. \ref{fig2}) and the bars take into account the standard error associated to this average.}
\label{fig3}
\end{figure} 

In Fig. \ref{fig3} we consider curves obtained averaging over $N=20$ measurements carried out with the same form of the envelop function $f(t)$. The first one (left panel) is achieved with a measurement time $t^{(1)}_{m}=600\,\,\mathrm{ns}$, while the second with $t^{(2)}_{m}=135\,\,\mathrm{ns}$, which is a value still consistent with the pulse constraints discussed above. Through best fit of the reported data according to Eq. (\ref{E_m_par}) one obtains 
\begin{eqnarray}
    a^{(1)}&=&0.981\pm 0.003\\
    \phi^{(1)}&=&0.45\pm 0.06
\end{eqnarray}
in the first case and 
\begin{eqnarray}
    a^{(2)}&=&0.964\pm 0.003\\
    \phi^{(2)}&=&0.30\pm 0.06
\end{eqnarray}
in the second. These data indicate that the Armonk qubit behaves as a QB with non zero, but limited, initial energy which can be almost completely charged with maximum stored energy $E^{(1/2)}_{max}$ exceeding $95\%$ in both cases. Moreover, the second case is characterized by a greater average charging power, namely the energy stored with respect to the charging time (which is necessary lower with respect to the measured time $t_{m}$)~\cite{Ferraro18, Andolina18}. In addition, the observed charging time and stored energy is comparable with recently reported state of the art measurements on the first charging step of a three level transmon QB~\cite{Hu21}. Considering even lower values of the measurement time, and consequently of the the standard deviation of the Gaussian pulse, data show a progressively stronger deviation from the predicted universal charging curve as a consequence of the emergence of the previously described technical limitations. 

It is worth noticing that the reported measurement (and consequently charging) times are substantially shorter with respect to both the decay ($T_{1}=165\,\,\mu\mathrm{s}$) and dephasing ($T_{2}=214\,\,\mu\mathrm{s}$) time reported for the machine. This represents a very strong indication of the fact that IBM quantum machines are already, without any ad hoc optimizations, in the proper range of parameters to be considered as quite efficient and stable QB.

To conclude this part we observe that the fluctuating phase acquired by the state at the level of the initialization, which is a major problem in a quantum computation perspective, can lead counter-intuitively to an improvement of the device as a QB. Indeed, the measured value of phase parameter $\phi$ is positive in the majority of the investigated cases. According to this, the value of $\theta$ at which the maximum charge is reached can occur for a value $\theta_{max}<\pi$, requiring a lower effective area of the envelop function to be achieved, with a consequent improvement of the performance of the QB. The effects of an imperfect initialization are even more evident by considering more general initial conditions which are discussed in the following.   

\subsection{More general initial conditions}

\begin{figure}[H]
\includegraphics[width=14.0 cm]{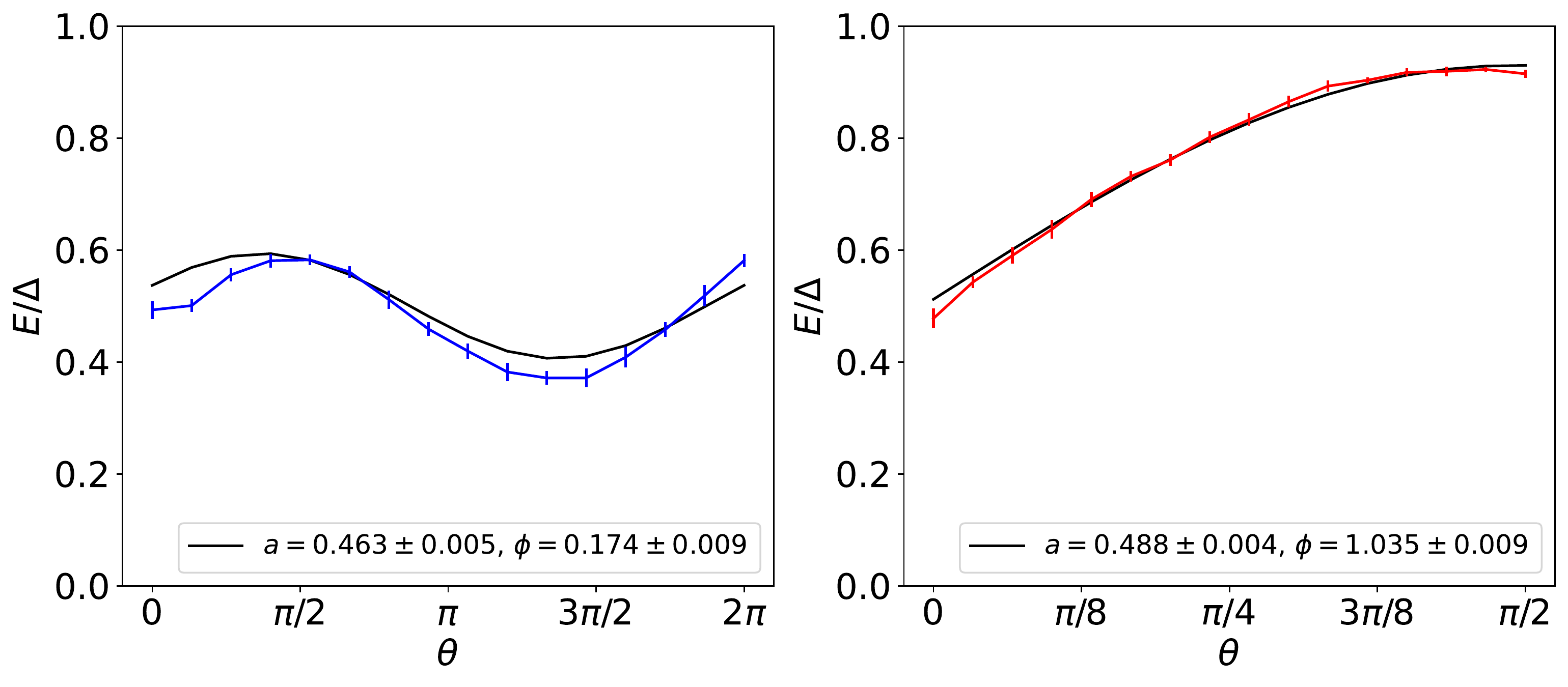}
\caption{(Color online). Best fit of the energy stored into the QB (in units of $\Delta$) as a function of $\theta$ (black curves). Data correspond to Gaussian pulses applied after the action of a unitary matrix $U$ (blue curve in the left panel) or $V$ (red curve in the right panels) over the ground state of the system. These transformations are defined in Eq. (\ref{U}) and Eq. (\ref{V}) of the main text respectively. Every point is obtained averaging over $10$ different measurements and the bars take into account the standard error associated to this average. In both cases the measurement time has been fixed at $t_{m}=600\,\,\mathrm{ns}$.}
\label{fig4}
\end{figure}

The initialization error discussed above can have an impact on the functioning of the QB also when more general quantum superposition states are considered as initial conditions. These states can be obtained starting from the ground state and applying a proper built-in unitary operator. As an example, in the following we will consider the two operators

\begin{equation}
U=\
\frac{1}{\sqrt{2}}\begin{pmatrix}
1 & 1 \\
1 & -1
 \end{pmatrix}
 \label{U}
\end{equation}

and 

\begin{equation}
V=\
\frac{1}{\sqrt{2}}\begin{pmatrix}
1 & +i \\
-i & -1
 \end{pmatrix}.
 \label{V}
\end{equation}
For an ideal QB the energy stored should be 
\begin{equation}
    E\left(\frac{1}{2}, 0, \theta\right)= \frac{\Delta}{2}
\end{equation}
in the first case, showing no charging/discharging dynamics due to the fact that the action of the drive only induce the rotation of this state on the equator of the Bloch sphere, and
\begin{equation}
    E\left(\frac{1}{2}, \frac{\pi}{2}, \theta\right)=\frac{\Delta}{2} \left(1+\sin \theta \right)
\end{equation}
in the second case, requiring half of the energy ($\Delta/2$) and half of the envelop area ($\theta/2$) to realize a complete charging. 

However, both the initialization and the application of a unitary operator are intrinsically affected by errors as can be seen from the corresponding measurements reported in Fig. \ref{fig4}. Again we observe that these errors can affect the functioning of the Armonk qubit as a QB. Indeed, even if limited, the first state (left panel) show an unexpected charging dynamics, while for the second state we observe a charging curve quite close to the ideal one, even if the maximum charging is not reached. Both these behaviors can be explained with the fact that the actual initial states are different with respect to what expected. In particular we have 
\begin{eqnarray}
a^{(U)}&=&0.463\pm 0.005\\
\phi^{(U)}&=& 0.174\pm 0.009
\end{eqnarray}
and 
\begin{eqnarray}
a^{(V)}&=&0.488\pm 0.004\\
\phi^{(V)}&=& 1.035\pm 0.009.
\end{eqnarray}
According to this, while the values of the amplitude are not very far for the ideal condition (few percent mismatch), the phase is again strongly affected by fluctuations.

\section{Conclusions}
\label{Conclusions}
In the present paper we have investigate the performances of the IBM Armonk single qubit in terms of energy storage and charging time by exploiting the Pulse tool included in the Qiskit package. This represents, as far as we know, the first actual simulation of a quantum battery using these kind of quantum devices.

With our analysis we have demonstrated that reasonably choosing a not very narrow and fast decaying classical drive it is possible to achieve very good energy storage (exceeding $95\%$) in a very short time (less then $135\,\,\mathrm{ns}$) with respect to the typical relaxation and dephasing time of the considered device. These performances are comparable with what observed in very recent state of the art experiments realized using superconducting qubits or semiconducting quantum dots. This indicates that the Armonk qubit, together with analogous machine in the same range of parameters, are already well designed to be seen as good and stable quantum batteries. 

Remarkably enough, their performances can be further improved by errors in the initialization state, such as phase fluctuations, which conversely have negative impact in the functioning of the device as qubit for quantum computation proposals. 

We think that this timely analysis could open new and fascinating perspectives in the fast developing field of quantum batteries and in the more general context of energy transfer devices addressing for example the controlled application of sequences of pulses in multi-qubit geometries.  
\vspace{6pt} 



\authorcontributions{Conceptualization, D.F. and M.G.; methodology, M.G.; software, G.G.; validation, G.G., and D.F.; formal analysis, G.G.; investigation, G.G.; data curation, G.G.; writing---original draft preparation, D.F.; writing---review and editing, G.G., M.G and M.S.; supervision, S.V. and M.S. All authors have read and agreed to the published version of the manuscript.}

\acknowledgments{G.G, D.F. and M.S. would like to acknowledge the support of "Dipartimento di Eccellenza MIUR 2018-22".
We acknowledge the use of IBM Quantum services for this work. The views expressed are those of the authors, and do not reflect the official policy or position of IBM or the IBM Quantum team.}

\conflictsofinterest{The authors declare no conflict of interest.} 

\sampleavailability{Data are available from the authors upon request.}


\abbreviations{Abbreviations}{
The following abbreviations are used in this manuscript:\\

\noindent 
\begin{tabular}{@{}ll}
QB & Quantum Battery\\
\end{tabular}}

\begin{adjustwidth}{-\extralength}{0cm}

\reftitle{References}

\end{adjustwidth}

\begin{thebibliography}{999}
%
\bibitem[Campaioli(2018)]{Campaioli18}
Campaioli, F.; Pollock, F. A.; Vinjanampathy, S. \textit{Thermodynamics in the Quantum Regime}, Springer, Berlin, 2018.
%
\bibitem[Bhattacharjee(2021)]{Bhattacharjee21}
Bhattacharjee, S.; Dutta, A. Quantum thermal machines and batteries. {\em Eur. Phys. J. B} {\bf 2021}, {\em 94}, 239.
%
\bibitem[Vincent(1997)]{Vincent_Book}
Vincent, C. A.; Scrosati, B. \textit{Modern Batteries}, Butterworth-Heinemann, Oxford, 1997.
%
\bibitem[Dell(2001)]{Dell_Book}
Dell, R. M.; Rand, D. A. J. \textit{Understanding Batteries}, The
Royal Society of Chemistry, Cambridge, 2001.
%
\bibitem[Alicki(2013)]{Alicki13}
Alicki, R.; Fannes, M. Entanglement boost for extractable work from ensembles of quantum batteries. {\em Phys. Rev. E} {\bf 2013}, {\em 87}, 042123.
%
\bibitem[Binder(2015)]{Binder15}
Binder, F. C.; Vinjanampathy, S.; Modi, K.; Goold, J. Quantacell: powerful charging of quantum batteries. {\em New J. Phys.} {\bf 2015}, {\em 17}, 075015.
%
\bibitem[Campaioli(2017)]{Campaioli17}
Campaioli, F.; Pollock, F. A.; Binder, F. C.; Céleri, L.; Goold, J.; Vinjanampathy, S.; Modi, K. Enhancing the Charging Power of Quantum Batteries. {\em Phys. Rev. Lett.} {\bf 2017}, {\em 118}, 150601.
%
\bibitem[JuliaFarre(2020)]{JuliaFarre20}
Julià-Farré, S.; Salamon, T.; Riera, A.; Bera, M. N.; Lewenstein, M. Bounds on the capacity and power of quantum batteries. {\em Phys. Rev. Research} {\bf 2020}, {\em 2}, 023113.
%
\bibitem[Gyhm(2022)]{Gyhm22}
Gyhm, J.-Y.; Safránek, D.; Rosa, D. Quantum Charging Advantage Cannot Be Extensive without Global Operations. {\em Phys. Rev. Lett.} {\bf 2022}, {\em 128}, 140501.
%
\bibitem[Le(2018)]{Le18}
Le, T. P.; Levinsen, J.; Modi, K.; Parish, M. M.; Pollock, F. A. 
Spin-chain model of a many-body quantum battery. {\em Phys. Rev. A} {\bf 2018}, {\em 97}, 022106.
%
\bibitem[Liu(2019)]{Liu19}
Liu, J.; Segal, D.; Hanna, G. A loss-free excitonic quantum battery. 
{\em J. Phys. Chem. C} {\bf 2019}, {\em 123}, 18303.
%
\bibitem[Rossini(2020)]{Rossini20}
Rossini, D.; Andolina, G. M.; Rosa, D.; Carrega, M.; Polini, M. 
Quantum Advantage in the Charging Process of Sachdev-Ye-Kitaev Batteries. {\em Phys. Rev. Lett.} {\bf 2020}, {\em 125}, 236402.
%
\bibitem[Rosa(2020)]{Rosa20}
Rosa, D.; Rossini, D.; Andolina, G. M.; Polini, M.; Carrega, M. Ultra-stable charging of fast-scrambling SYK quantum batteries. {\em J. High Energ. Phys.} {\bf 2020}, {\em 67}, 2020.
%
\bibitem[Crescente(2020)]{Crescente20}
Crescente, A.; Carrega, M.; Sassetti, M.; Ferraro, D. Charging and energy fluctuations of a driven quantum battery. {\em New J. Phys.} {\bf 2020}, {\em 22}, 063057.
%
\bibitem[Carrega(2020)]{Carrega20}
Carrega, M.; Crescente, A.; Ferraro, D.; Sassetti, M. Dissipative dynamics of an open quantum battery. {\em New J. Phys.} {\bf 2020}, {\em 22}, 083085.
%
\bibitem[Santos(2021)]{Santos21}
Santos, A.C. Quantum advantage of a two-level batteries in self-discharging process. {\em Phys. Rev. E} {\bf 2021}, {\em 103}, 042118.
%
\bibitem[Peng(2021)]{Peng21}
Peng, L.; He, W.-B.; Chesi,S.; Lin, H.-Q.; Guan, X.-W. Lower and upper bounds of quantum battery power in multiple central spin systems. {\em Phys. Rev. A} {\bf 2021}, {\em 103}, 052220.
%
\bibitem[Ferraro(2018)]{Ferraro18}
Ferraro, D.; Campisi, M.; Andolina, G. M.; Pellegrini, V.; Polini, M. High-Power Collective Charging of a Solid-State Quantum Battery. {\em Phys. Rev. Lett.} {\bf 2018}, {\em 120}, 117702.
%
\bibitem[Ferraro(2019)]{Ferraro19}
Ferraro, D.; Andolina, G. M.; Campisi, M.; Pellegrini, V.; Polini, M. Quantum supercapacitors. {\em Phys. Rev. B} {\bf 2019}, {\em 100}, 075433.
%
\bibitem[Crescente(2020b)]{Crescente20b}
Crescente, A.; Carrega, M.; Sassetti, M.; Ferraro, D. Ultrafast charging in a two-photon Dicke quantum battery. {\em Phys. Rev. B} {\bf 2020}, {\em 102}, 245407.
%
\bibitem[Delmonte(2021)]{Delmonte21}
Delmonte, A.; Crescente, A.; Carrega, M.; Ferraro, D.; Sassetti, M. Characterization of a Two-Photon Quantum Battery: Initial Conditions, Stability and Work Extraction. {\em Entropy} {\bf 2021}, {\em 23}, 612.
%
\bibitem[Dou(2022)]{Dou22}
Dou, F.-Q.; Lu, Y.-Q.; Wang, Y.-J.; Sun, J.-A. Extended Dicke quantum battery with interatomic interactions and driving field.
{\em Phys. Rev. B} {\bf 2022}, {\em 105}, 115405.
%
\bibitem[Quanch(2022)]{Quach22}
Quach, J. Q.; McGhee, K. E.; Ganzer, L.; Rouse, D. M.; Lovett, B. W.; Gauger, E. M.; Keeling, J.; Cerullo, G.; Lidzey, D. G.; Virgili, T. 
Superabsorption in an organic microcavity: Toward a quantum battery.
{\em Science Advances} {\bf 2022}, {\em 8}, eabk3160.
%
\bibitem[Hu(2021)]{Hu21}
Hu, C.-K.; Qiu, J.; Souza, P. J. P.; Yuan, J.; Zhou, Y.; Zhang, L.; Chu, J.; Pan, X.; Hu, L.; Li, J.; Xu, Y.; Zhong, Y.; Liu, S.; Yan, F.; Tan, D.; Bachelard, R.; Villas-Boas, C. J.; Santos, A. C.; Yu, D. Optimal charging of a superconducting quantum battery. {\em arXiv} {\bf 2021}, {\em arXiv:2108.04298}.
%
\bibitem[Wenniger(2022)]{Wenniger22}
Maillette de Buy Wenniger, I.; Thomas, S. E.; Maffei, M.; Wein, S. C.; Pont, M.; Harouri, A.; Lemaitre, A.; Sagnes, I.; Somaschi, N.; Auffèves, A.; Senellart, P. Coherence-powered work exchanges between a solid-state qubit and light fields. {\em arXiv} {\bf 2022}, {\em arXiv:2202.01109}.
%
\bibitem[Andolina(2018)]{Andolina18}
Andolina, G. M.; Farina, D.; Mari, A.; Pellegrini, V.; Giovannetti, V.; Polini, M. Charger-mediated energy transfer in exactly solvable models for quantum batteries. {\em Phys. Rev. B} {\bf 2018} {\em 98}, 205423.
%
\bibitem[Qi(2021)]{Qi21}
Qi, S.-F.; Jing, J. Magnon-mediated quantum battery under systematic errors. {\em Phys. Rev. A} {\bf 2021}, {\em 104}, 032606.
%
\bibitem[Crescente(2022)]{Crescente22}
Crescente, A.; Ferraro, D.; Carrega, M.; Sassetti, M. Enhancing coherent energy transfer between quantum devices via a mediator. {\em arXiv} {\bf 2022}, {\em arXiv:2202.01025}.
%
\bibitem[Zhang(2019)]{Zhang19}
Zhang, Y.-Y.; Yang, T.-R.; Fu, L.; Wang, X. Powerful harmonic charging in a quantum battery. {\em Phys. Rev. E} {\bf 2019}, {\em 99}, 052106.
%
\bibitem[Chen(2020)]{Chen20}
Chen, J.; Zhan, L.; Shao, L.; Zhang, X.; Zhang, Y.-Y.; Wang, X. Charging Quantum Batteries with a General Harmonic Driving
Field. {\em Ann. Physik} {\bf 2020}, {\em 532}, 1900487.
%
\bibitem[Corcoles(2020)]{Corcoles20}
Córcoles, A. D.; Kandala, A.; Javadi-Abhari, A.; McClure, D. T.; Cross, A. W.; Temme, K.; Nation, P. D.; Steffen, M.; Gambetta, J. M.
Challenges and Opportunities of Near-Term Quantum Computing Systems. 
{\em Proceedings of the IEEE} {\bf 2020}, {\em 108}, 1338.
%
\bibitem[Cao(2019)]{Cao_2019}
Cao, Y.; Romero, J.; Olson, J. P.; Degroote, M.; Johnson, P. D.; Kieferov{\'{a}}, M.; Kivlichan, I. D.; Menke, T.; Peropadre, B.;  Sawaya, N. P. D.; Sim, S.; Veis, L.; Aspuru-Guzik, A.
Quantum Chemistry in the Age of Quantum Computing.
{\em Chemical Reviews} {\bf 2019}, {\em 119}, 19.
%
\bibitem[Guimar(2020)]{Guimar_2020}
Guimaraes, J. D.; Tavares, C.; Soares, L.; Vasilevskiy; Mikhail I.
Simulation of Nonradiative Energy Transfer in Photosynthetic Systems Using a Quantum Computer.
{\em Complexity} {\bf 2020}, {\em 2020}, 1.
%
\bibitem[Chiesa(2019)]{Chiesa_2019}
Chiesa, A.; Tacchino, F.; Grossi, M.; Santini, P.; Tavernelli, I.; Gerace, D.; Carretta, S.
Quantum hardware simulating four-dimensional inelastic neutron scattering.
{\em Nature Physics} {\bf 2019}, {\em 15}, 5.
%
\bibitem[Crippa(2021)]{Crippa_2021}
Crippa, L.; Tacchino, F.; Chizzini, M.; Aita, A.; Grossi, M.; Chiesa, A.; Santini, P.; Tavernelli, I.; Carretta, S.
Simulating Static and Dynamic Properties of Magnetic Molecules with Prototype Quantum Computers.
{\em Magnetochemistry} {\bf 2021}, {\em 7}, 8.
%
\bibitem[FillionGourdeau(2017)]{Fillion_Gourdeau_2017}
Fillion-Gourdeau, F.; MacLean, S.; Laflamme, R.
Algorithm for the solution of the Dirac equation on digital quantum computers.
{\em Physical Review A} {\bf 2017}, {\em 95}, 4.
%
\bibitem[Klco(2020)]{Klco_2020}
Klco, N.; Savage, M. J.; Stryker, J. R.
SU(2) non-Abelian gauge field theory in one dimension on digital quantum computers.
{\em Physical Review D} {\bf 2020}, {\em 101}, 7.
%
\bibitem[Bauer(2021)]{bauer2021quantum}
Nachman, B.; Provasoli, D.; de Jong, W. A.; Bauer, C. W.
Quantum Algorithm for High Energy Physics Simulations.
{\em Phys. Rev. Lett.} {\bf 2021}, {\em 126}, 6.
%
\bibitem[Grossi(2022)]{grossi_2022}
Agliardi, G.; Grossi, M.; Pellen, M.; Prati, E.
Quantum integration of elementary particle processes.
{\em arXiv} {\bf 2022}, {\em 2201.01547}.
%
\bibitem[Cervia(2020)]{Cervia2020ExactlySM}
Cervia, J. C.; Balantekin, A. B.; Coppersmith, S. N.; Johnson, C. W.; Love, P.J.; Poole, C.; Robbins, K.; Saffman, M.
Exactly solvable model as a testbed for quantum-enhanced dark matter detection.
{\em arXiv} {\bf 2020}, {\em 2201.01547}.
%
\bibitem[Pistoia(2022)]{Pistoia22}
Herman, D.; Googin, C.; Liu, X.; Galda, A.; Safro, I.; Sun, Y.; Pistoia, M.; Alexeev, Y.
A Survey of Quantum Computing for Finance.
{\em arXiv} {\bf 2022}, {\em 2201.02773}.
%
\bibitem[Gao(2021)]{Gao2021}
Gao, Q.; Jones, G. O.; Motta, M.; Sugawara, M.; Watanabe, H. C.; Kobayashi, T.; Watanabe, E.; Ohnishi, Y.;  Nakamura, H.; Yamamoto, N.
Applications of quantum computing for investigations of electronic transitions in phenylsulfonyl-carbazole {TADF} emitters.
{\em npj Computational Materials} {\bf 2021}, {\em 7}, 1.
%
\bibitem[Moll(2018)]{Moll_2018}
Moll, N.; Barkoutsos, P.; Bishop, L. S.; Chow, J. M.; Cross, A.; Egger, D. J.; Filipp, S.; Fuhrer, A.; Gambetta, J. M., Ganzhorn, M. et al.
Quantum optimization using variational algorithms on near-term quantum devices.
{\em Quantum Sci. Technol.} {\bf 2018}, {\em 3}, 3.
%
\bibitem[Alexander(2020)]{Alexander20}
Alexander, T.; Kanazawa, N.; Egger, D. J.; Capelluto, L.; Wood, C. J.; Javadi-Abhari, A.; McKay, D. C. Qiskit pulse: programming quantum computers through the cloud with pulses. {\em Quantum Sci. Technol.} {\bf 2020} {\em 5}, 044006.
%
\bibitem[Preskill(2018)]{Preskill18}
Preskill, J. Quantum Computing in the NISQ era and beyond. {\em Quantum} {\bf 2018}, {\em 2}, 79.
%
\bibitem[Koch(2007)]{Koch07}
Koch, J.; Yu, T. M.; Gambetta, J.; Houck, A. A.; Schuster, D. I.; Majer, J.; Blais, A.; Devoret, M. H.; Girvin, S. M.; Schoelkopf, R. J. Charge-insensitive qubit design derived from the Cooper pair box. {\em Phys. Rev. A} {\bf 2007}, {\em 76}, 042319.
%

\bibitem[Schleich(2021)]{Schleich_Book}
Schleich, W. P. \textit{Quantum Optics in Phase Space}, Wiley VCH, Berlin, 2021.
%


\bibitem[Krantz(2019)]{Krantz19}
Krantz, P.; Kjaergaard, M.; Yan, F.; Orlando, T. P.; Gustavsson, S.; Oliver, W. D. A Quantum Engineer’s Guide to Superconducting Qubits. {\em Appl. Phys. Rev.} {\bf 2019}, {\em 6}, 021318.
%
\bibitem[Jeffrey(2014)]{Jeffrey14}
Jeffrey, E.; Sank, D.; Mutus, J. Y.; White, T. C.; Kelly, J.; Barends, R.; Chen, Y.; Chen, Z.; Chiaro, B.; Dunsworth, A.; Megrant, A.; O'Malley, P. J. J.; Neill, C.; Roushan, P.; Vainsencher, A.; Wenner, J.; Cleland, A. N.; Martinis, J. M. Fast Accurate State Measurement with Superconducting Qubits. {\em Phys. Rev. Lett.} {\bf 2014}, {\em 112}, 190504.
%

\end{thebibliography}
\end{document}